%% file: ms.tex
\documentclass[nobibnotes,floatfix,aps,twocolumn,preprintnumbers]{revtex4-1}
\usepackage{amsmath,amssymb,bm,float,Macro,mathrsfs}
\usepackage[dvips]{graphicx}
\usepackage{color}

\input{local-macro}

\begin{document}

%$$$$$$$$$$$$$$$$$$$$$$$$$$$$$$$$$$$$$$$$$$$$$$$$$$$$$$$$$$$$$$$$$$$$$$$$$$$$$$$$$$$$$$$$$$$$$$$$$$$$%
%////////////////////////////////////////////////////////////////////////////////////////////////////%
% FRONT MATTER 
%////////////////////////////////////////////////////////////////////////////////////////////////////%
%$$$$$$$$$$$$$$$$$$$$$$$$$$$$$$$$$$$$$$$$$$$$$$$$$$$$$$$$$$$$$$$$$$$$$$$$$$$$$$$$$$$$$$$$$$$$$$$$$$$$%

% Title %
\title{Constraining cosmic string parameters with curl mode of CMB lensing} 

% Authors %
% ---- Toshiya Namikawa ---- 
\author{Toshiya Namikawa}
%\email{namikawa@yukawa.kyoto-u.ac.jp}
\affiliation{Yukawa Institute for Theoretical Physics, Kyoto University, Kyoto 606-8502, Japan}
% ---- Daisuke Yamauchi ---- 
\author{Daisuke Yamauchi}
%\email{yamauchi@resceu.s.u-tokyo.ac.jp}
\affiliation{Research Center for the Early Universe, School of Science, The University of Tokyo, Bunkyo-ku, Tokyo 113-0033, Japan}
% ---- Atsushi Taruya ---- 
\author{Atsushi Taruya}
%\email{ataruya@utap.phys.s.u-tokyo.ac.jp}
\affiliation{Yukawa Institute for Theoretical Physics, Kyoto University, Kyoto 606-8502, Japan}
\affiliation{Research Center for the Early Universe, School of Science, The University of Tokyo, Bunkyo-ku, Tokyo 113-0033, Japan}
\affiliation{Institute for the Physics and Mathematics of the Universe, The University of Tokyo, Kashiwa, Chiba 277-8568, Japan}

% Date %
\date{\today}

\preprint{RESCEU-39/13,\,YITP-13-83}

% Abstract %
%----------------------------------------------------------------------------------------------------%
\begin{abstract}
We present constraints on a cosmic string network with a measurement of weak gravitational lensing 
from CMB temperature map. 
The cosmic string network between observer and last scattering surface of CMB photons 
generates vector and/or tensor metric perturbations, and the deflection of CMB photons by these 
gravitational fields has curl mode which is not produced by the scalar metric perturbations. 
In this paper, we use the power spectrum of curl mode obtained from 
Planck to constrain the string tension, $G\mu$, and the reconnection probability, $P$. 
In demonstrating the parameter constraints with Planck curl mode, 
we also measure the lensing power spectrum from the Atacama Cosmology Telescope (ACT) 2008 season data, 
which have better angular resolution with lower instrumental noise on a much smaller chunk of the sky. 
Assuming $P=1$, the upper bound on tension is $G\mu =6.6\times 10^{-5}$ with $2\sigma$ \,(95\% C.L.), 
using curl mode from Planck, which is weaker than that from the small-scale temperature power spectrum. 
For small values of $P$, however, the constraint from curl mode becomes tighter compared to that from 
temperature power spectrum. For $P\lsim 10^{-2}$, we obtain the constraint on the combination of the 
string parameters as $G\mu P^{-1}\leq 3.4\times 10^{-5}$ at more than $2\sigma$ \,(95\% C.L.).
\end{abstract} 
%----------------------------------------------------------------------------------------------------%

\maketitle

%////////////////////////////////////////////////////////////////////////////////////////////////////%
% MAIN MATTER 
%////////////////////////////////////////////////////////////////////////////////////////////////////%

% Contents %
\input{ms-sec1}

\input{ms-sec2}
\input{ms-sec3}

\input{ms-sec4}

\input{ms-sec5}

\input{ms-sec6}

%////////////////////////////////////////////////////////////////////////////////////////////////////%
% BACK MATTER 
%////////////////////////////////////////////////////////////////////////////////////////////////////%

% Acknowledgments %
\begin{acknowledgments}
We greatly appreciate the Planck team for kindly providing us with the curl-mode power spectrum 
obtained from Planck temperature maps. 
We also thank Duncan Hanson for helpful discussion and comments, and the anonymous referee for 
improving the text. 
% Grants 
DY and AT are supported in part by a Grants-in-Aid for Scientific Research from the Japan Society for 
the Promotion of Science (JSPS) (No. 259800 for DY and No. 24540257 for AT). 
This work was supported in part by Grant-in-Aid for Scientific Research on Priority Areas No. 467 
``Probing the Dark Energy through an Extremely Wide and Deep Survey with Subaru Telescope''. 
% Computers
Part of numerical computations were carried out at the Yukawa Institute Computer Facility. 
\end{acknowledgments}

% Appendix %
\appendix
\input{ms-appA}

\input{ms-appB}

% References %
\bibliographystyle{mybst}
\bibliography{cite}

\end{document}

%% file: local-macro.tex
\def\tT{\widetilde{\Theta}}
\def\hT{\widehat{\Theta}}

\def\hCTT{\widehat{C}^{\Theta\Theta}}
\def\tCTT{\widetilde{C}^{\Theta\Theta}}
\def\hC{\widehat{C}}

\def\hx{\widehat{x}}

\def\grad{\phi}
\def\curl{\varpi}

\def\Ag{\alpha^{\mC{E}}}
\def\Ac{\alpha^{\mC{B}}}

\def\hAg{\widehat{\alpha}^{\mC{E}}}

\def\bl{\bm{\ell}}
\def\bL{\bm{L}}
\def\hatn{\widehat{\bm{n}}}
\def\bn{\bm{\nabla}}

%% file: ms-sec1.tex
%****************************************************************************************************%
%////////////////////////////////////////////////////////////////////////////////////////////////////%
\section{\label{sec1}Introduction}
%////////////////////////////////////////////////////////////////////////////////////////////////////%
%****************************************************************************************************%

%<><><><><><><><><><><><><><><><><><><><><><><><><><><><><><><><><><><><><><><><><><><><><><><><><><>%
\begin{figure*}[t]
\bc
\vspace*{-5\baselineskip}
\hspace*{5em}
\includegraphics[width=180mm,clip]{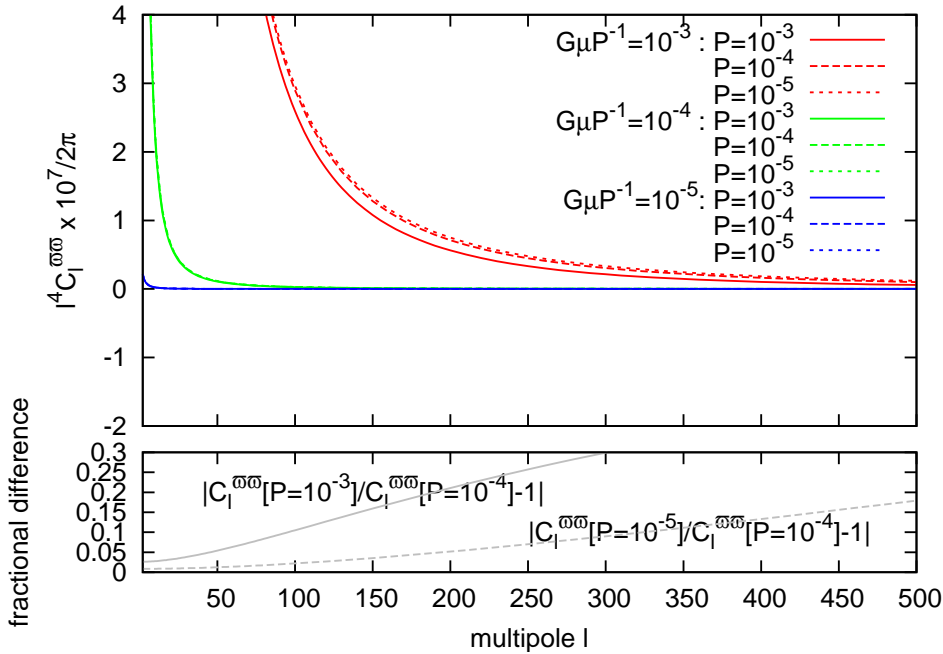} 
\caption{
Theoretical angular power spectra of curl mode for different values of the combination of 
the cosmic string parameters with $G\mu P^{-1}=10^{-3}$ (red), $10^{-4}$ (green), $10^{-5}$ (blue). 
For each value of $G\mu P^{-1}$, we show the case with different values of the reconnection 
probability, i.e., $P=10^{-3}$ (solid), $10^{-4}$ (dashed), $10^{-5}$ (dotted). 
The bottom panel shows the fractional difference between the angular power spectra with
the different values of $P$.
}
\label{Fig:thCl}
\ec
\end{figure*}
%<><><><><><><><><><><><><><><><><><><><><><><><><><><><><><><><><><><><><><><><><><><><><><><><><><>%

% Introduction (Recent measurements and future prospects of CMB lensing) 
At arcminute scales, CMB anisotropies are perturbed significantly by gravitational lensing. 
Recently, several groups have reported the detection of lensing signals by reconstructing lensing 
fields involved in the CMB anisotropies, using CMB maps alone or the cross correlations between CMB 
and other observables \citep{Smith07,Hirata:2008cb,Das:2011ak,vanEngelen:2012va,Bleem:2012gm,
Sherwin:2012mr,Das:2013zf,Ade:2013tyw,Ade:2013aro,Hanson:2013daa}. 
The lensing signals measured from ongoing, upcoming and next-generation CMB experiments, e.g., 
Planck \cite{PLANCK}, SPTpol \cite{SPTPol}, PolarBear \cite{POLARBEAR}, ACTPol \cite{Niemack:2010wz}, 
COrE \cite{Bouchet:2011ck}, PRISM \cite{Andre:2013afa} and CMBPol \cite{Smith:2008an}, 
will have enough sensitivity to probe several fundamental issues, such as properties of dark energy 
and massive neutrinos (e.g., \cite{Hu:2001fb,Lesgourgues:2006nd,dP09,Namikawa:2010re,Das:2011mf}). 

% Lensing effect on CMB and introduction to gradient and curl modes
The lensing effect on the temperature anisotropies is described by a remapping of the primary 
anisotropies. Denoting the primary temperature anisotropies at position $\hatn$ on the last scattering 
surface, $\Theta(\hatn)$, the lensed temperature anisotropies are given by 
%----------------------------------------------------------------------------------------------------%
\al{
	\tT(\hatn) 
		&= \Theta(\hatn+\bm{d}(\hatn)) \notag \\ 
		&= \Theta(\hatn) + \bm{d}(\hatn)\cdot\bm{\nabla}\Theta(\hatn) + \mC{O}(|\bm{d}|^2)
	\,. \label{remap}
}
%----------------------------------------------------------------------------------------------------%
The vector, $\bm{d}(\hatn)$, is the deflection angle, and, in terms of parity, we can decompose it 
into gradient (even parity) and curl (odd parity) modes 
\cite{Hirata:2003ka,Cooray:2005hm,Namikawa:2011cs}: 
%----------------------------------------------------------------------------------------------------%
\al{
	\bm{d}(\hatn) = \bn \grad(\hatn) + (\star\bn)\curl(\hatn)
	\,, 
}
%----------------------------------------------------------------------------------------------------%
where the symbol, $\star$, denotes an operation which rotates the angle of a two-dimensional vector 
counterclockwise by $90$ degrees. 

% Motivation 
It is known that scalar metric perturbations such as the matter density fluctuations at linear 
order produce the gradient mode, but do not generate the curl mode. 
At linear order, the curl mode is induced by only vector and/or tensor metric perturbations.
Thus, the curl mode is a probe of nonscalar metric fluctuations. 

In this paper, among various active seeds of nonscalar metric perturbations, we are particularly 
interested in cosmic strings. They can be generally formed as topological defects in the early 
Universe through a symmetry breaking phase transition. There has been a renewed interest in another 
possibility that F and D strings of superstring theory can appear at the end of stringy inflation and 
act as cosmic strings, so-called cosmic superstrings. The evolution of cosmic superstrings may differ 
from that of ordinary field-theoretic strings, because the reconnection probability $P$ may be 
significantly smaller than unity. While the ordinary strings always reconnect when they intersect, 
the reconnection probability of cosmic superstrings is typically in the range 
$10^{-3}\lesssim P\lesssim 1$. Moreover, $P$ strongly depends on the details of the compactification.
For a hybrid network that contains bound states, the situation is more complicated and $P$ can vary 
from $0$ to $1$~\cite{Copeland:2003bj,Jackson:2004zg}. 
Hence, the constraint on $P$ plays an important role in distinguishing between ordinary cosmic strings 
and cosmic superstrings.

According to our previous work \cite{Namikawa:2011cs,Yamauchi:2012bc,Yamauchi:2013fra}, 
cosmic strings can produce an observable curl mode for small values of $P$. 
Detection of a curl mode in the CMB lensing deflection would provide strong evidence for 
cosmic strings. An interesting point is that the lensing curl mode is more sensitive to small values 
of the reconnection probability compared to the temperature angular power spectrum induced by 
the Gott-Kaiser-Stebbins (GKS) effect \cite{Kaiser:1984,Gott:1985}. 
These two probes from CMB observation would have complementarity for constraining cosmic string 
parameters \cite{Yamauchi:2013prep}. In this paper, using the angular power spectrum of the curl 
modes, we present constraints on cosmic string parameters. 
In our analysis of parameter constraints, we use the curl-mode power 
spectrum from Planck. In demonstrating the results obtained from Planck data, 
we also measure the lensing power spectrum from the ACT data, which have better angular 
resolution with lower instrumental noise on a much smaller chunk of the sky. 

% Structure of the paper
This paper is organized as follows. 
In Sec.~\ref{sec2}, we briefly review a cosmic string network and its effect on 
the curl-mode power spectrum. 
In Sec.~\ref{sec3}, our curl-mode data for constraining cosmic string parameters are described, 
and the method for obtaining our curl-mode power spectrum is explained. 
In Sec.~\ref{sec4}, we show the curl-mode power spectra used in our analysis, and 
in Sec.~\ref{sec5}, the results of constraints on cosmic-string parameters are shown. 
Sec.~\ref{sec6} is devoted to summary and discussion of future prospects.

%% file: ms-sec2.tex
%****************************************************************************************************%
%////////////////////////////////////////////////////////////////////////////////////////////////////%
\section{Curl mode from cosmic string network} \label{sec2} 
%////////////////////////////////////////////////////////////////////////////////////////////////////%
%****************************************************************************************************%

%Extended objects such as 
Cosmic strings continuously produce not only scalar perturbations but also vector and tensor 
perturbations, and induce curl modes \cite{Namikawa:2011cs,Yamauchi:2012bc} 
(and also B-mode shear \cite{Yamauchi:2012bc} and rotation \cite{Thomas:2009bm} in galaxy images). 
The amplitudes of the curl modes are typically determined by the dimensionless string tension $G\mu$\,. 
In this paper, we focus on the contributions from vector perturbations since those from tensor 
perturbations are negligible in the curl mode for the scales of our interest~\cite{Yamauchi:2013fra}. 

The curl-mode angular power spectrum induced by vector perturbations in general has the following 
form \cite{Yamauchi:2012bc,Yamauchi:2013fra}: 
%----------------------------------------------------------------------------------------------------%
\al{ 
	&C_\ell^{\varpi\varpi}
		= 4\pi\frac{(\ell -1)!}{(\ell +1)!}\int^\infty_0\frac{{\rm d}k}{k} 
			\biggl[
				\int^{\chi_*}_0\frac{{\rm d}\chi}{\chi}\Delta_1 (k,\chi )
				j_\ell (k\chi )
			\biggr]^2
	\,,\label{eq:curl angular power spectrum}
} 
%----------------------------------------------------------------------------------------------------%
where $\chi_*$ denotes the conformal distance at the last scattering surface and 
$\Delta_1 (k,\chi )$ is the dimensionless auto power spectrum of the vector perturbations.
Note that the curl-mode power spectrum given above has originally involved the unequal-time power 
spectrum. Here, to evaluate the integral analytically, we adopt the factorizable ansatz that 
the unequal-time power spectrum is described by the square root of the product of auto power spectra 
given at different times. 

To compute the string-induced power spectrum, we consider a string network described by
the velocity-dependent one-scale model (VOS)~\cite{Martins:1996jp,Martins:2000cs,Avgoustidis:2005nv,Takahashi:2008ui,Yamauchi:2010ms,Yamauchi:2010vy},
which is characterized by the correlation length $\xi =1/H\gamma$ and the root-mean-square velocity 
$v$\,. Here $H$ denotes the Hubble expansion rate.
In the VOS, the reconnection process is required for the energy loss of the network through loop 
formation and the approach to the scaling solution, in which the correlation length scales with 
the Hubble radius. Thus, observables associated with the global properties of the network sensitively 
depend on the reconnection probability, and $\gamma$, $v$ in the scaling regime are in fact well 
approximated by $\gamma\approx (\pi\sqrt{2}/3\tilde cP)^{1/2}$\,, $v^2\approx (1/2)(1-\pi/(3\gamma))$ 
for $\tilde cP\ll 1$\,, where $\tilde c\approx 0.23$ quantifies the efficiency of the loop formation~
\cite{Martins:2000cs}.
If $G\mu$ and $P$ are given, the quantities, $\xi$ and $v$, are evaluated, and then the power spectrum 
induced by cosmic strings is computed as 
%----------------------------------------------------------------------------------------------------%
\al{ 
	\Delta_1^2 (k,\chi )
		&= (16\pi G\mu )^2
			\frac{\sqrt{6\pi}v^2}{12(1-v^2)} 
	\notag \\ 
	&\qquad \times
			\frac{4\pi k^3\chi^2 a^4}{H}
			\left(\frac{a}{k\xi}\right)^5
			{\rm erf}\left(\frac{k\xi /a}{2\sqrt{6}}\right)
	\,.\label{eq:vector power spectrum}
} 
%----------------------------------------------------------------------------------------------------%
The above equation implies that the angular power spectrum given in 
Eq.~\eqref{eq:curl angular power spectrum} roughly scales as 
$C_{\ell}^{\curl\curl}\propto (G\mu)^2 P^{-2}$\, at the large-scale limit ($\ell\to 0$), 
where the curl-mode signals become significant. 

In Fig.~\ref{Fig:thCl}, to see the dependence of curl mode on $G\mu$ and $P$, 
we show the expected angular power spectrum for the curl mode with a combination 
of parameters, $G\mu P^{-1}=10^{-3}$ (red), $10^{-4}$ (green), $10^{-5}$ (blue). 
For each case of $G\mu P^{-1}$, we change the values of the reconnection probability 
as $P=10^{-3}$ (solid), $10^{-4}$ (dashed), and $10^{-5}$ (dotted). 
The bottom panel shows the fractional difference between the angular power spectra with the different 
values of $P$. 
The string-induced curl mode has large amplitude at large-angular scales. 
As expected, for the same value of $G\mu P^{-1}$, the curl-mode power spectra at large scales are 
hardly distinguishable. 
Hence, measurement of the curl mode gives constraints on the combination 
of string parameters $G\mu P^{-1}$\,. 

%% file: ms-sec3.tex
%****************************************************************************************************%
%////////////////////////////////////////////////////////////////////////////////////////////////////%
\section{\label{sec3}Data and Analysis} 
%////////////////////////////////////////////////////////////////////////////////////////////////////%
%****************************************************************************************************%

To constrain cosmic string parameters, we use the curl-mode power spectrum measured from two 
independent data sets. 
The first set of curl-mode data we use in this paper is obtained from Planck \cite{Ade:2013tyw}, 
a satellite CMB experiment. 
The curl mode obtained from Planck has the highest signal-to-noise ratio in the currently available 
CMB data. 
The resultant curl-mode power spectrum from Planck, however, may be affected by, e.g., 
residual contaminations from foregrounds and analysis of lensing reconstruction. 
As a second data set, we derive a curl-mode power spectrum estimate from a public ACT temperature 
map, and the details are described below. 
Since ACT has high angular resolution and sensitive to temperature fluctuations at smaller scales 
compared to Planck, the systematics involved in the ACT data would be different from that in the 
Planck data.
In this respect, the results obtained from the ACT data can be utilized as a cross-check of 
the Planck results.
The measured curl mode obtained from ACT is useful to understand how the constraints on cosmic string 
parameters depend on the sensitivity of CMB experiments to the curl-mode signal. 
In Secs.~\ref{sec3.1} and \ref{sec3.2}, we briefly summarize the method for estimating the curl-mode 
power spectrum from the ACT and Planck temperature maps, respectively.

%****************************************************************************************************%
%::::::::::::::::::::::::::::::::::::::::::::::::::::::::::::::::::::::::::::::::::::::::::::::::::::%
\subsection{ACT} \label{sec3.1}
%::::::::::::::::::::::::::::::::::::::::::::::::::::::::::::::::::::::::::::::::::::::::::::::::::::%
%****************************************************************************************************%

%****************************************************************************************************%
%::::::::::::::::::::::::::::::::::::::::::::::::::::::::::::::::::::::::::::::::::::::::::::::::::::%
\subsubsection{Temperature map} 
%::::::::::::::::::::::::::::::::::::::::::::::::::::::::::::::::::::::::::::::::::::::::::::::::::::%
%****************************************************************************************************%

For lensing reconstruction, we use the full map with point source subtraction, taken from the LAMBDA 
website \cite{LAMBDA}, observed at 148 GHz in the 2008 season, covering 845.6 deg$^2$ of the sky 
\cite{Dunner:2012vp}. 
Choosing a region where the noise level is lower than $\sim 35\,\mu$K-arcmin, the full map is divided 
into four rectangular regions as follows. 
The coordinate of the center for $i$th map ($i=1$ - $4$) is given by $(1500 + 1920\times i, 588)$ in 
grid space. 
The area of each quarter map is $16\times 4$ deg$^2$ (where the size of each pixel is $0.5$ 
arcmin-square). 
To mitigate survey boundary effects, we apply an apodization window function following 
Ref.~\cite{Namikawa:2012pe}: 
%----------------------------------------------------------------------------------------------------%
\al{
	W(x,y;s_0)=w(x;s_0)w(y;s_0) 
	\,. 
}
%----------------------------------------------------------------------------------------------------%
We use a sine apodization function given by
%----------------------------------------------------------------------------------------------------%
\al{
	w(s;s_0) &= w_0\times 
		\begin{cases} 1 & |s|<as_0 \\ 
			\sin\left(\dfrac{\pi}{2}\dfrac{1-|s|/a}{1-s_0}\right) & as_0\leq |s|<a \\ 
			0 & a\leq |s| 
		\end{cases}
	\,, \label{apo}
}
%----------------------------------------------------------------------------------------------------%
with $w_0=(2a[s_0+2(1-s_0)/\pi])^{-1}$. 
Note that the parameter, $s_0$, indicates the width of the region where the apodization is applied. 
Throughout this paper, we choose $s_0=0.0$.

%****************************************************************************************************%
%::::::::::::::::::::::::::::::::::::::::::::::::::::::::::::::::::::::::::::::::::::::::::::::::::::%
\subsubsection{Lensing potential estimator} 
%::::::::::::::::::::::::::::::::::::::::::::::::::::::::::::::::::::::::::::::::::::::::::::::::::::%
%****************************************************************************************************%

The estimator for the lensing potentials, $\hx_{\bl}$ ($x=\grad$ or $\curl$), can be constructed 
using the statistical anisotropy generated by lensing
(see e.g., Refs.~\cite{Zaldarriaga:1998te,Hu:2001kj} for gradient mode, and 
Refs.~\cite{Cooray:2005hm,Namikawa:2011cs} for curl mode). 
A naive estimator, however, suffers from a so-called ``mean-field bias,'' i.e., nonzero mean of 
the estimator in the presence of masking, inhomogeneous noise and so on. 
The mean-field bias needs to be corrected with some methods. 

In our analysis, to reduce the mean-field bias, we use the following estimator \cite{Namikawa:2012pe}: 
%----------------------------------------------------------------------------------------------------%
\al{
	\hx_{\bL} & = \mC{A}_{L}^{xx}\Int{2}{\bl}{(2\pi)^2}
		\frac{\mC{F}^x_{\bL,\bl}}{2\hCTT_{\ell}\hCTT_{|\bL-\bl|}}\hT_{\bl}\hT_{\bL-\bl} 
	\,. \label{Eq:est}
}
%----------------------------------------------------------------------------------------------------%
The quantity, $\hT_{\bl}$, is the beam-deconvolved observed map in Fourier space and expressed as 
$\hT_{\bl} = \tT_{\bl} + B_{\ell}^{-1}n_{\bl}$, with $B_{\ell}$ denoting the isotropic beam transfer 
function, taken from the LAMBDA website \cite{LAMBDA}, and $n_{\bl}$ describing the instrumental noise 
as well as contaminations from unresolved point sources. 
The quantity $\hCTT_{\ell}$ is the temperature angular power spectrum measured directly from each map. 
The weight function and normalization are defined as 
%----------------------------------------------------------------------------------------------------%
\al{
	\mC{F}_{\bL,\bl}^{x} &= 
		\sum_{a=x,\epsilon,s} \frac{A_{L}^{aa}\{\bR{R}_{L}^{-1}\}^{xa}}
		{A_{L}^{xx}\{\bR{R}_{L}^{-1}\}^{xx}}F_{\bL,\bl}^{a} 
		\,, \\ 
	\mC{A}_{L}^{xx} &= \{\bR{R}_{L}^{-1}\}^{xx}A_{L}^{xx} 
		\label{Eq:mcA}
	\,. 
}
%----------------------------------------------------------------------------------------------------%
Note that the indexes, $\epsilon$ and $s$, represent the effect of inhomogeneous reionization, 
Doppler boosting, and of additional contaminations such as inhomogeneous noise and unresolved point 
sources, respectively (see Ref.~\cite{Namikawa:2012pe} for details). 
The quantity, $\bR{R}_{L}$, is the $3\times 3$ response matrix whose elements are 
%----------------------------------------------------------------------------------------------------%
\al{
	R_{L}^{ab} = \frac{A_{L}^{aa}}{A_{L}^{ab}} \,; \,
	A_{L}^{ab} = \left\{\Int{2}{\bl}{(2\pi)^2} \frac{F_{\bL,\bl}^{a}F_{\bL,\bl}^{b}}
		{2\hCTT_{\ell}\hCTT_{|\bL-\bl|}}\right\}^{-1}
	\,, \label{eqn:responsefunc}
}
%----------------------------------------------------------------------------------------------------%
where the weight functions are given by 
%----------------------------------------------------------------------------------------------------%
\al{
	F^{\grad}_{\bL,\bl} &= [\tCTT_{\ell}\bL\cdot\bl + \tCTT_{|\bL-\bl|}\bL\cdot(\bL-\bl)] \notag \\ 
	F^{\curl}_{\bL,\bl} &= [\tCTT_{\ell}(\star\bL)\cdot\bl + \tCTT_{|\bL-\bl|}(\star\bL)\cdot(\bL-\bl)]
		\notag \\ 
	F^{\epsilon}_{\bL,\bl} &= - \hCTT_{\ell} - \hCTT_{|\bL-\bl|} \,, \notag \\ 
	F^{s}_{\bL,\bl} &= 1  
	\,, \label{sec2:weight}
}
%----------------------------------------------------------------------------------------------------%
with $\delta_{\bm{0}}\tCTT_{\ell}=\ave{|\tT_{\bl}|^2}$ denoting the lensed temperature angular power 
spectrum, computed by {\tt CAMB} \cite{Lewis:1999bs} with fiducial cosmological parameters from 
ACT+WMAP data \cite{Dunkley:2010ge}. 
Note that, for curl mode, since $R^{\curl\epsilon}_{\ell}=R^{\curl s}_{\ell}=0$, we find 
$\mC{F}_{\bl,\bL}^{\varpi}=F_{\bl,\bL}^{\varpi}$ and 
$\mC{A}_{\ell}^{\varpi\varpi}=A_{\ell}^{\varpi\varpi}$.

%****************************************************************************************************%
%::::::::::::::::::::::::::::::::::::::::::::::::::::::::::::::::::::::::::::::::::::::::::::::::::::%
\subsubsection{Lensing power spectrum estimate: Bias-hardened estimator} 
\label{sec:BHE}
%::::::::::::::::::::::::::::::::::::::::::::::::::::::::::::::::::::::::::::::::::::::::::::::::::::%
%****************************************************************************************************%

With the estimators for lensing fields described above, the lensing power spectrum, 
$\hC^{xx}_{L}$, is estimated as \cite{Namikawa:2012pe,Ade:2013tyw} 
%----------------------------------------------------------------------------------------------------%
\al{
	\hC^{xx}_{L} = \INT{}{\varphi_{\bL}}{2\pi}{0}{2\pi} [|\hx_{\bL}|^2 - \widehat{N}^{xx}_{\bL} ] 
	\,, \label{Eq:est-Clxx}
}
%----------------------------------------------------------------------------------------------------%
with $\varphi_{\bL}$ denoting the angle of multipole vector, $\bL$. 
Hereafter, we call the above estimator the bias-hardened estimator (BHE). 
The second term in the above equation (\ref{Eq:est-Clxx}), usually referred to as 
the ``Gaussian bias,'' is estimated through \cite{Namikawa:2012pe}
%----------------------------------------------------------------------------------------------------%
\al{
	\widehat{N}^{xx}_{\bL} 
		&= \left( \mC{A}_{L}^{xx} \right)^2 
		\Int{2}{\bl}{(2\pi)^2}\Int{2}{\bl'}{(2\pi)^2} 
	\notag \\ 
		&\times \frac{\mC{F}^x_{\bL,\bl}\mC{F}^x_{\bL,\bl'}}
		{\hCTT_{\ell}\hCTT_{|\bL-\bl|}\hCTT_{\ell'}\hCTT_{|\bL-\bl'|}}
	\notag \\ 
		&\times \left(\bR{C}_{\bl,\bL-\bl'}\hT_{\bL-\bl}\hT_{\bl'} 
		- \frac{1}{2}\bR{C}_{\bl,\bL-\bl'}\bR{C}_{\bL-\bl,\bl'} \right)
	\,, \label{Eq:GN}
}
%----------------------------------------------------------------------------------------------------%
where $\delta_{\bl-\bl'}\bR{C}_{\bl,\bl'}\!=\!\ave{\hT_{\bl}\hT_{\bl'}}$ is the theoretical ensemble 
of covariance matrix for observed multipoles. 

%****************************************************************************************************%
%::::::::::::::::::::::::::::::::::::::::::::::::::::::::::::::::::::::::::::::::::::::::::::::::::::%
\subsubsection{Lensing power spectrum estimate: $\ell$-splitting}
\label{sec:LSP}
%::::::::::::::::::::::::::::::::::::::::::::::::::::::::::::::::::::::::::::::::::::::::::::::::::::%
%****************************************************************************************************%

The lensing power spectrum estimator given in Eq.~(\ref{Eq:est-Clxx}) usually suffers from several 
uncertainties. The estimator for the lensing power spectrum has the Gaussian bias term (\ref{Eq:GN}), 
which usually has large contributions. The contribution from first order of $C_{\ell}^{xx}$, usually 
referred to as N1 bias \cite{Kesden:2003cc,BenoitLevy:2013bc}, also causes a non-negligible 
contamination in the estimation of the lensing power spectrum on small scales. 
A convenient approach to mitigate these biases is to use the $\ell$-splitting method (LSP)
\cite{Hu:2001fa,Sherwin:2010ge}. 
In this method, the temperature multipoles are divided into two disjoint annular regions. 
The estimated cross-power spectrum between these two reconstructed maps has no Gaussian and is 
insensitive to N1 biases, although the signal-to-noise ratio for the lensing power spectrum decreases 
compared to the usual technique. In our analysis, we perform lensing reconstruction not only with the 
realization-dependent Gaussian bias subtraction [see Eq.~\eqref{Eq:GN}] but also with 
$\ell$-splitting, and confirm that the results obtained from the two methods are consistent with 
each other.

%****************************************************************************************************%
%::::::::::::::::::::::::::::::::::::::::::::::::::::::::::::::::::::::::::::::::::::::::::::::::::::%
\subsection{Planck} \label{sec3.2}
%::::::::::::::::::::::::::::::::::::::::::::::::::::::::::::::::::::::::::::::::::::::::::::::::::::%
%****************************************************************************************************%

In our analysis, the published curl-mode power spectrum of Ref.~\cite{Ade:2013tyw} is used. 
In the Planck lensing analysis \cite{Ade:2013tyw}, the temperature maps measured at 143 and 217 GHz
are used for their main results of lensing reconstruction, reducing contaminations from Galactic 
foregrounds, carbon-monoxide emission lines (for 217 GHz) and point sources by masking. 
In their analysis of lensing reconstruction, they use the quadratic estimator  
for estimating gradient and curl modes based on Refs.~\cite{Okamoto:2003zw} and 
\cite{Namikawa:2011cs}, respectively, and then the mean-field biases are subtracted with Monte Carlo 
simulations. 
Note that, for gradient mode, as a cross-check, 
they also use the mean-field reduced estimator (\ref{Eq:est}) in full-sky lensing reconstruction. 
The lensing power spectrum is then estimated based on Eq.(\ref{Eq:GN}) in the full sky case. 
They also compute the lensing power spectrum obtained by combining 143 and 217 GHz results, which 
is used in this paper.

%% file: ms-sec4.tex
%****************************************************************************************************%
%////////////////////////////////////////////////////////////////////////////////////////////////////%
\section{Lensing power spectra} 
\label{sec4} 
%////////////////////////////////////////////////////////////////////////////////////////////////////%
%****************************************************************************************************%

%<><><><><><><><><><><><><><><><><><><><><><><><><><><><><><><><><><><><><><><><><><><><><><><><><><>%
\begin{figure*}[t]
\bc
\includegraphics[width=89mm,clip]{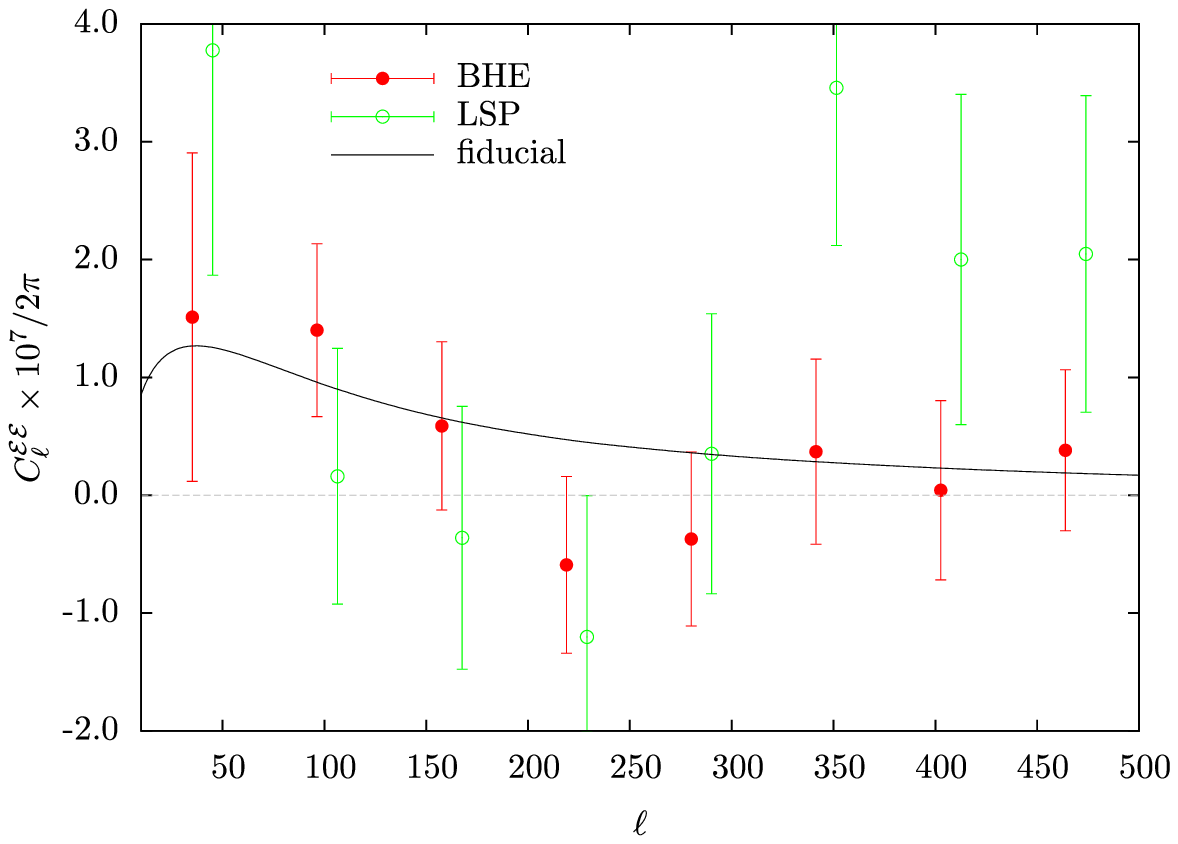} 
\includegraphics[width=89mm,clip]{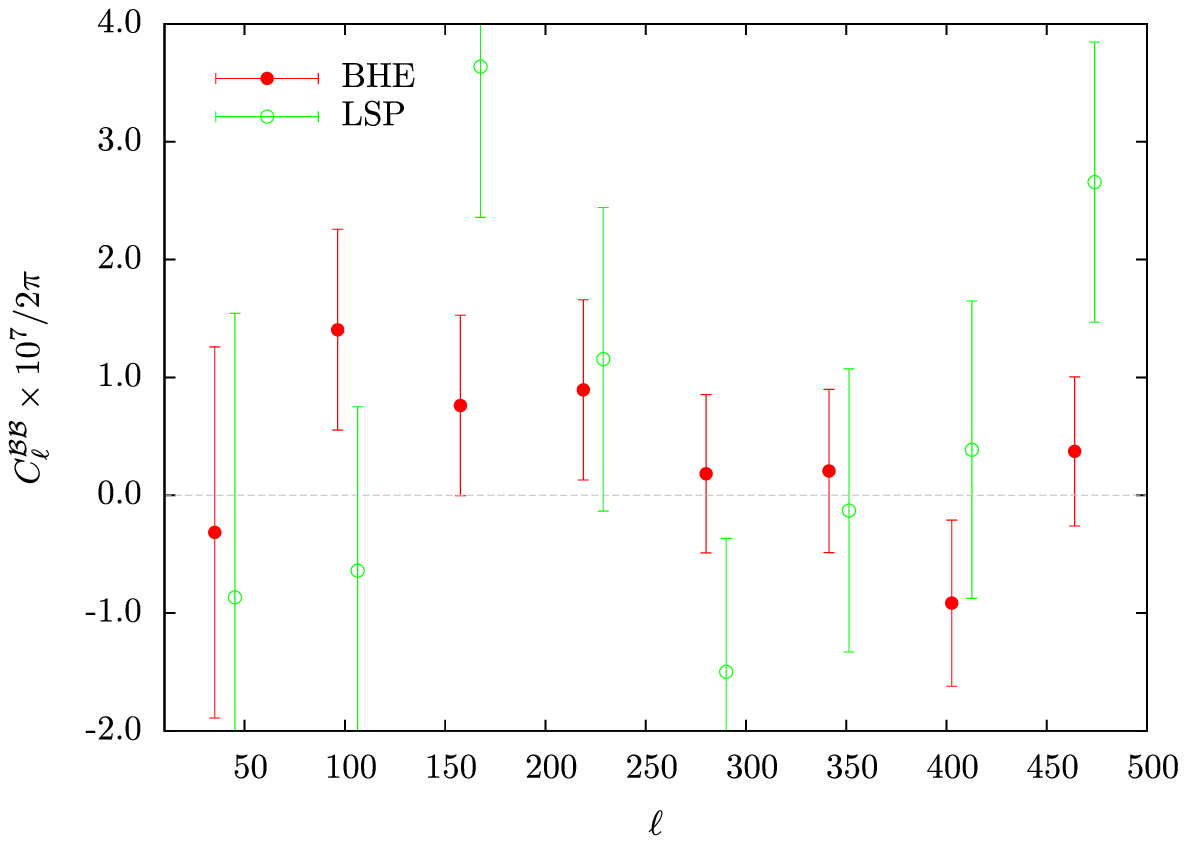}
\caption{
{\it Left}: The angular power spectrum for gradient modes obtained with the $\ell$-splitting 
method (LSP; green), compared with our fiducial method, based on Eq.~(\ref{Eq:GN}) (BHE; red). 
The expected power spectrum with fiducial parameters from \cite{Dunkley:2010ge}
is shown as a black dashed line.
{\it Right}: Same as right panel, but for curl mode. 
}
\label{Fig:comp}
\ec
\end{figure*}
%<><><><><><><><><><><><><><><><><><><><><><><><><><><><><><><><><><><><><><><><><><><><><><><><><><>%

%<><><><><><><><><><><><><><><><><><><><><><><><><><><><><><><><><><><><><><><><><><><><><><><><><><>%
\begin{figure*}[t]
\bc
\includegraphics[width=89mm,clip]{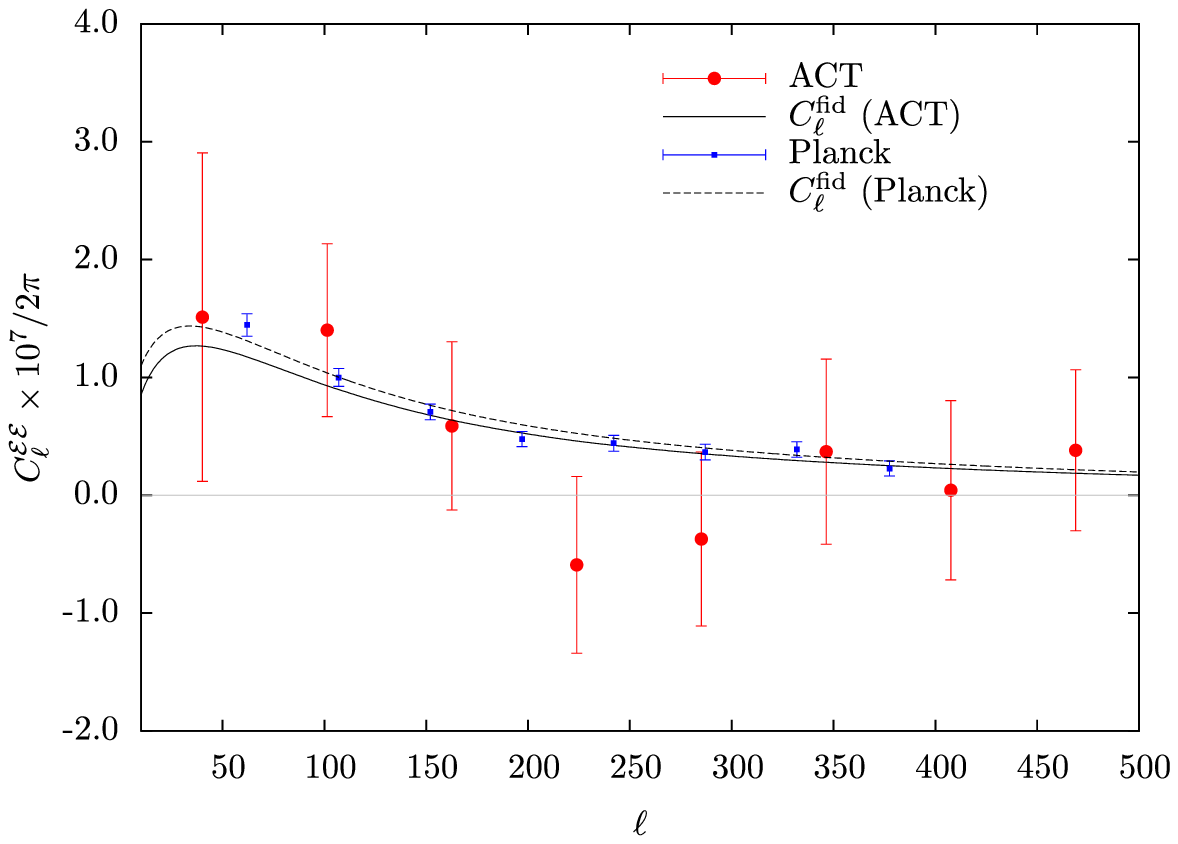} 
\includegraphics[width=89mm,clip]{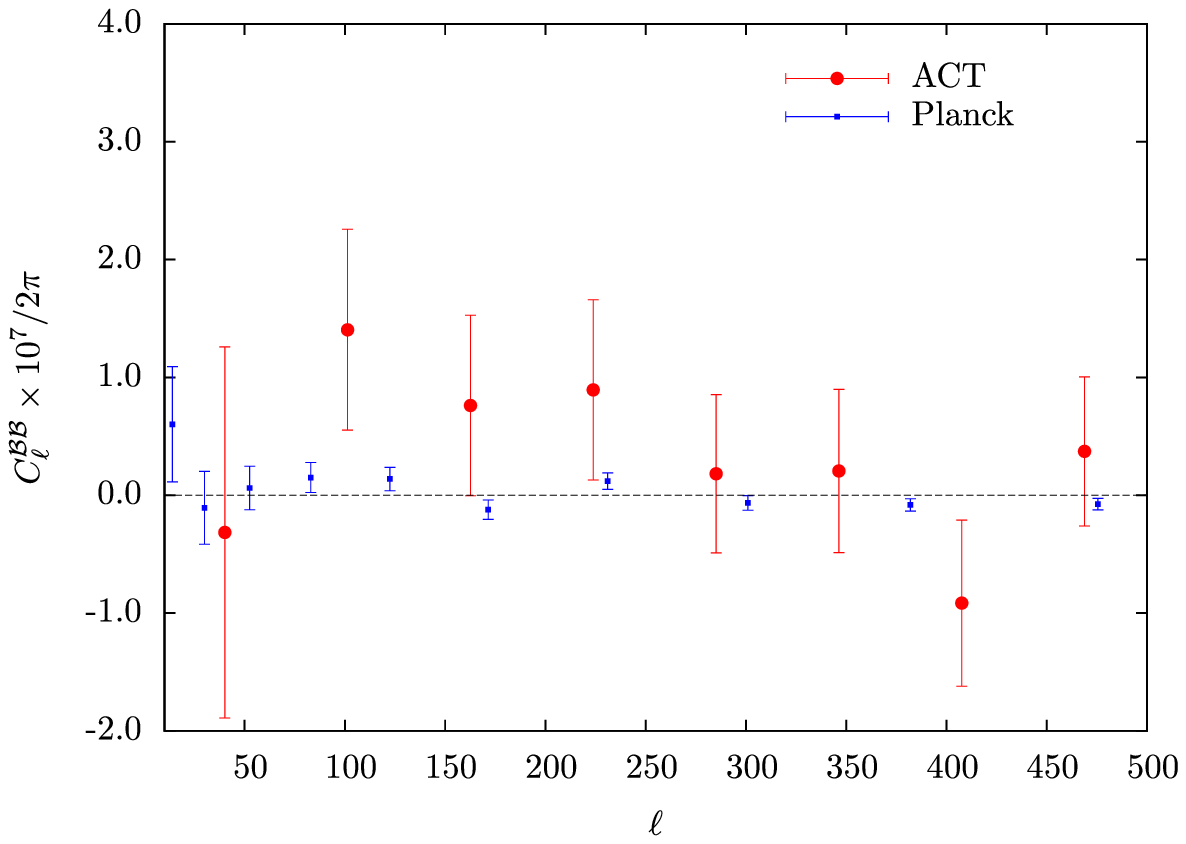}
\caption{
{\it Left}: Estimated angular power spectrum for gradient mode obtained from ACT (red) and Planck 
(blue), compared with the expected power spectrum with fiducial parameters from 
Ref.~\cite{Dunkley:2010ge} (ACT; black solid) and Ref.~\cite{Ade:2013tyw} (Planck; black dashed). 
{\it Right}: Same as right panel, but for curl mode. 
}
\label{Fig:ACT-measured}
\ec
\end{figure*}
%<><><><><><><><><><><><><><><><><><><><><><><><><><><><><><><><><><><><><><><><><><><><><><><><><><>%

%<><><><><><><><><><><><><><><><><><><><><><><><><><><><><><><><><><><><><><><><><><><><><><><><><><>%
\begin{table}[t]
\caption{
\label{Tab:constraints}
Comparison of parameter constraints on $\Ag$ and $\Ac$ with different methods and data. 
}
\vs{0.5}
\begin{tabular}{c|cccc} \hline 
 & BHE & LSP & ACT \cite{Das:2011ak} & Planck \\ \hline 
$\Ag$ & $0.91\pm 0.46$ & $1.06\pm 0.66$ & $1.16\pm 0.29$ & $0.94\pm 0.04$ \cite{Ade:2013tyw} \\ \hline 
$\Ac$ & $<0.80$ & $<0.97$ & $-$ & $<0.014$ \\ \hline 
\end{tabular}
\end{table}
%<><><><><><><><><><><><><><><><><><><><><><><><><><><><><><><><><><><><><><><><><><><><><><><><><><>%

In this section, we show the angular power spectra of gradient and curl modes defined as 
%----------------------------------------------------------------------------------------------------%
\al{
	C_L^{\mC{E}\mC{E}}\equiv L^4C_{L}^{\grad\grad} 
	\,, \qquad 
	C_L^{\mC{B}\mC{B}}\equiv L^4C_{L}^{\curl\curl} 
	\,. 
}
%----------------------------------------------------------------------------------------------------%
To compare with Planck results, we compute the binned angular power spectra following the Planck 
analysis described in Ref.~\cite{Ade:2013tyw}. The details are shown in the following. 

%****************************************************************************************************%
%::::::::::::::::::::::::::::::::::::::::::::::::::::::::::::::::::::::::::::::::::::::::::::::::::::%
\subsection{Multipole binning} \label{sec4.1}
%::::::::::::::::::::::::::::::::::::::::::::::::::::::::::::::::::::::::::::::::::::::::::::::::::::%
%****************************************************************************************************%

For gradient mode, to compute the binned angular power spectrum, we first estimate the amplitude of 
the lensing power spectrum defined as $\hAg_L=\hC_L^{\mC{E}\mC{E}}/C_L^{\mC{E}\mC{E},\rm{fid}}$, where 
$\hC_{L}^{\mC{E}\mC{E}}$ and $C_{L}^{\mC{E}\mC{E},{\rm fid}}$ denote the measured and expected power 
spectra, respectively. 
The expected lensing power spectrum, $C_L^{\mC{E}\mC{E},{\rm fid}}$, is computed from the best-fit 
values of the temperature power spectrum from \cite{Dunkley:2010ge} (\cite{Ade:2013zuv}) for ACT 
(Planck). 
Note that, by constraining $\hAg_L$, we can show whether the 
measured lensing power spectrum is consistent with the lensing 
power spectrum expected from the flat $\Lambda$ cold dark matter framework. 
Then, the amplitude parameter at the $b$th multipole bin, $\hAg_b$, is estimated as \cite{Ade:2013tyw} 
%----------------------------------------------------------------------------------------------------%
\al{
	\hAg_{b} \equiv (\sigma^{\mC{E}}_b)^2\sum_{L^b\rom{min}\leq L <L^b\rom{max}} \mS{B}^{\mC{E}}_L\hAg_L
	\,, \label{Eq:Ab} 
}
%----------------------------------------------------------------------------------------------------%
where $L^b\rom{min}$ and $L^b\rom{max}$ are the minimum and maximum multipoles of the $b$th bin, and 
the band-pass function and the variance of $\hAg_b$ for the $b$th bin are given by 
%----------------------------------------------------------------------------------------------------%
\al{
	\mS{B}^{\mC{E}}_L = \left(\frac{C_L^{\mC{E}\mC{E},{\rm fid}}}{\Delta C_L^{\mC{E}\mC{E}}}\right)^2 
	\,, \qquad 
	\sigma^{\mC{E}}_b = \left\{\sum_L \mS{B}^{\mC{E}}_{L}\right\}^{-1/2} 
	\,. \label{Eq:bandpass-error} 
}
%----------------------------------------------------------------------------------------------------%
The error on angular power spectrum, $\Delta C_L^{\mC{E}\mC{E}}$ [and also $\Delta C_L^{\mC{B}\mC{B}}$ 
appearing later in Eq.~(\ref{Eq:Ac})], is estimated from Eqs.~\eqref{Eq:est-error} and 
\eqref{Eq:est-error-lsp} in Appendix \ref{appA}, respectively, for the bias-hardened estimator and 
$\ell$-splitting methods. The method to estimate the map-combined power spectrum is also described in 
Appendix \ref{appA}. For binning in multipoles, we choose $L^b\rom{min}=10+(b-1)\times 490/n$ and 
$L^b\rom{max}=10+b\times 490/n$, where the number of multipole bins is $n=8$. 
The measured power spectrum in the $b$th mulipole bin is then obtained by scaling $\hAg_b$ with 
the expected power spectrum as 
%----------------------------------------------------------------------------------------------------%
\al{
	\hC^{\mC{E}\mC{E}}_b \equiv \hAg_b C^{\mC{E}\mC{E},{\rm fid}}_{L_b} 
	\,, 
}
%----------------------------------------------------------------------------------------------------%
with $L_b=(L^b\rom{min}+L^b\rom{max})/2$. The error bars are also multiplied by 
$C^{\mC{E}\mC{E},{\rm fid}}_{L_b}$. On the other hand, for the curl modes, we compute the measured 
power spectrum in the $b$th bin as 
%----------------------------------------------------------------------------------------------------%
\al{
	\hC^{\mC{B}\mC{B}}_b \equiv (\sigma^{\mC{B}}_b)^2\sum_{L^b\rom{min}\leq L <L^b\rom{max}} 
		\frac{1}{(\Delta C_{L}^{\mC{B}\mC{B}})^2} \hC^{\mC{B}\mC{B}}_L
	\,, \label{Eq:Ac} 
}
%----------------------------------------------------------------------------------------------------%
where $\sigma^{\mC{B}}_b=[\sum_L(\Delta C_L^{\mC{B}\mC{B}})^{-2}]^{-1/2}$. 

%****************************************************************************************************%
%::::::::::::::::::::::::::::::::::::::::::::::::::::::::::::::::::::::::::::::::::::::::::::::::::::%
\subsection{Statistical analysis} \label{sec.4-2}
%::::::::::::::::::::::::::::::::::::::::::::::::::::::::::::::::::::::::::::::::::::::::::::::::::::%
%****************************************************************************************************%

Let us first discuss whether our measured gradient and curl modes are statistically consistent with 
other results. Provided $\hAg_b$, the total lensing amplitude, $\Ag$, is estimated by minimizing 
%----------------------------------------------------------------------------------------------------%
\al{
	-2\ln \mC{L} (\Ag) = \sum_{b=1}^{n} \frac{(\hAg_b - \Ag)^2}{(\sigma^{\mC{E}}_b)^2}
	\,. \label{Eq:chi-sq}
}
%----------------------------------------------------------------------------------------------------%
The lensing power spectrum is estimated from both the BHE and LSP. 
The temperature multipoles with $500\leq \ell\leq 2000$ are used for BHE, 
and with the two disjoint annuli, $500\leq \ell\leq 1500$ and $1600\leq\ell\leq 2300$, for LSP. 
For curl mode, we estimate a parameter, $\Ac$, by minimizing the following likelihood: 
%----------------------------------------------------------------------------------------------------%
\al{
	-2\ln \mC{L} (\Ac) = \sum_{b=1}^n \frac{(\hC^{\mC{B}\mC{B}}_b-\Ac C^{\mC{B}\mC{B},{\rm fid}}_b)^2}
		{(\sigma^{\mC{B}}_b)^2}
	\,, \label{Eq:chi-sq-c}
}
%----------------------------------------------------------------------------------------------------%
where the fiducial binned power spectrum, $C^{\mC{B}\mC{B},{\rm fid}}_b$, is computed with the same 
binning as the measured power spectrum, using the theoretical power spectrum with 
$G\mu=1.3\times 10^{-9}$ and $P=4.5\times 10^{-6}$ which is chosen so that the resultant $1\sigma$ 
upper bound on $\Ac$ with ACT data roughly becomes unity. 

Now we show the results of constraints on $\Ag$ and $\Ac$, which are summarized in Table 
\ref{Tab:constraints}. 
For the gradient mode, with our measured power spectrum from ACT, the constraint on parameter $\Ag$ 
becomes $\Ag=0.91 \pm 0.46$ ($1\sigma$, BHE) and $\Ag=1.06 \pm 0.66$ ($1\sigma$, LSP),which is 
consistent with the Planck result within $1\sigma$ statistical significance, i.e., 
$\Ag=0.943\pm 0.040$ ($1\sigma$). 
Our result is also consistent with $\Ag=1.16\pm 0.29$ ($1\sigma$) obtained from the ACT map 
with noise level $\simeq 23 \,\mu$K presented in Ref.~\cite{Das:2011ak}. 
Note that the degradation of statistical significance of our $\Ag$ constraint compared to 
Ref.~\cite{Das:2011ak} would be due to the noise level and the use of the estimator, 
Eq.~(\ref{Eq:est}), for reducing mean-field bias. 

On the other hand, for curl mode, we find $\Ac < 0.80$ ($1\sigma$, BHE)
and $\Ac <0.97$ ($1\sigma$, LSP) for our measured curl mode from ACT, and 
$\Ac < 0.014$ $(1\sigma)$ for Planck curl mode. 
Our analysis shows that we only obtain the upper bound for curl mode, and the curl mode is 
consistent with zero. 
For ACT data, further discussions on several systematics are given in Appendix \ref{appB}. 

%****************************************************************************************************%
%::::::::::::::::::::::::::::::::::::::::::::::::::::::::::::::::::::::::::::::::::::::::::::::::::::%
\subsection{Binned angular power spectra} 
%::::::::::::::::::::::::::::::::::::::::::::::::::::::::::::::::::::::::::::::::::::::::::::::::::::%
%****************************************************************************************************%

In Fig.~\ref{Fig:comp}, we show the measured power spectrum for gradient (left) and curl modes (right) 
obtained by both BHE and LSP. 
For gradient modes, the measured power spectrum is compared with $C_L^{\mC{E}\mC{E},{\rm fid}}$. 
The mean of the measured power spectrum at several bins has a discrepancy with significance greater 
than $1\sigma$ level between the cases with BHE and LSP, both for gradient and curl modes. 
There are several possible systematics which would cause the discrepancy in our analysis; if there 
exist some additional sources of generating non-Gaussianity in the observed anisotropies, the 
normalization $A_L$ given in Eq.~(\ref{eqn:responsefunc}) which is derived in an idealistic case would 
lead to some amount of bias in the lensing estimator. 
The estimation of statistical error of the lensing power spectrum is also affected by the incorrect 
normalization, and underestimation of statistical errors in each bin causes the statistical 
discrepancy at each multipole bin. 
Moreover, the discrepancy is also caused by the systematics on LSP, in which we assume that the 
temperature multipoles on the two disjoint annuli are correlated solely due to the lensing effect. 
Nevertheless, the estimated total lensing amplitudes, $\Ag$ and $\Ac$, obtained from two different 
methods are consistent within $1\sigma$ level. 
In Appendix \ref{appB}, we also check the robustness of our analysis by changing multipole ranges, 
number of bins and size of multipole bins, and we confirm that the constraints on $\Ag$ and $\Ac$ 
obtained from these analyses are not so different from the constraints shown in Sec.~\ref{sec.4-2}. 
Since the amplitude of the string-induced curl-mode power spectrum is determined by the combination of 
$G\mu$ and $P$, the constraints on $G\mu$ and $P$ are almost determined by the constraints on the 
amplitude of curl-mode power spectrum, $\Ac$. 
Thus, we expect that the impact of the residual systematics in the curl-mode power spectrum is not so 
significant on the final results of the parameter constraints.
The statistical consistency between BHE and LSP would also indicate that the uncertainties in the 
subtraction of Gaussian bias and the contamination of N1 bias are also negligible. 
In the subsequent analysis, we use the BHE to constrain the string parameters. 

In Fig.~\ref{Fig:ACT-measured}, the comparison of lensing power spectra obtained from ACT and Planck 
are shown. For gradient modes, the measured power spectrum agrees well with that obtained from Planck, 
and has a similar trend of multipole dependence. 
The resultant estimates of $\Ag$ are also consistent with each other within $1\sigma$ level. 
For curl modes, although the multipole binning is different between Planck and our results, the 
resultant constraints on $\Ac$ obtained from our results and Planck are consistent. 

Before closing this section, we comment on the systematic effect of foreground contaminations on 
the curl-mode estimation. 
As mentioned in Ref.~\cite{Das:2011ak}, the lensing reconstruction with temperature maps around 148 GHz 
suffers from the foregrounds such as IR point sources and the Sunyaev-Zel'dovich (SZ) effect. 
Following Ref.~\cite{Das:2011ak}, we filter out the temperature multipoles below $500$ to mitigate 
atmospheric effect, and $\ell\gsim2000$ to avoid foreground point sources. 
In the analysis of Planck \cite{Ade:2013aro}, they use the temperature multipoles up to $\ell=1700$, 
and compare the resultant curl-mode power spectrum from 143 GHz with that from 217 GHz. 
In the above, we find that the resultant constraints on $\Ac$ are both consistent with zero for Planck 
and ACT. 
Since the constraints on string parameters basically come from those on $\Ac$, we expect that the 
residual foreground contaminations and the systematics in the Gaussian bias subtraction will not 
seriously affect the constraints on cosmic string parameters in both Planck and ACT cases.

%% file: ms-sec5.tex
%****************************************************************************************************%
%////////////////////////////////////////////////////////////////////////////////////////////////////%
\section{Constraints on cosmic string parameters} 
\label{sec5} 
%////////////////////////////////////////////////////////////////////////////////////////////////////%
%****************************************************************************************************%

%<><><><><><><><><><><><><><><><><><><><><><><><><><><><><><><><><><><><><><><><><><><><><><><><><><>%
\begin{figure*}[t]
\bc
\includegraphics[width=120mm,clip]{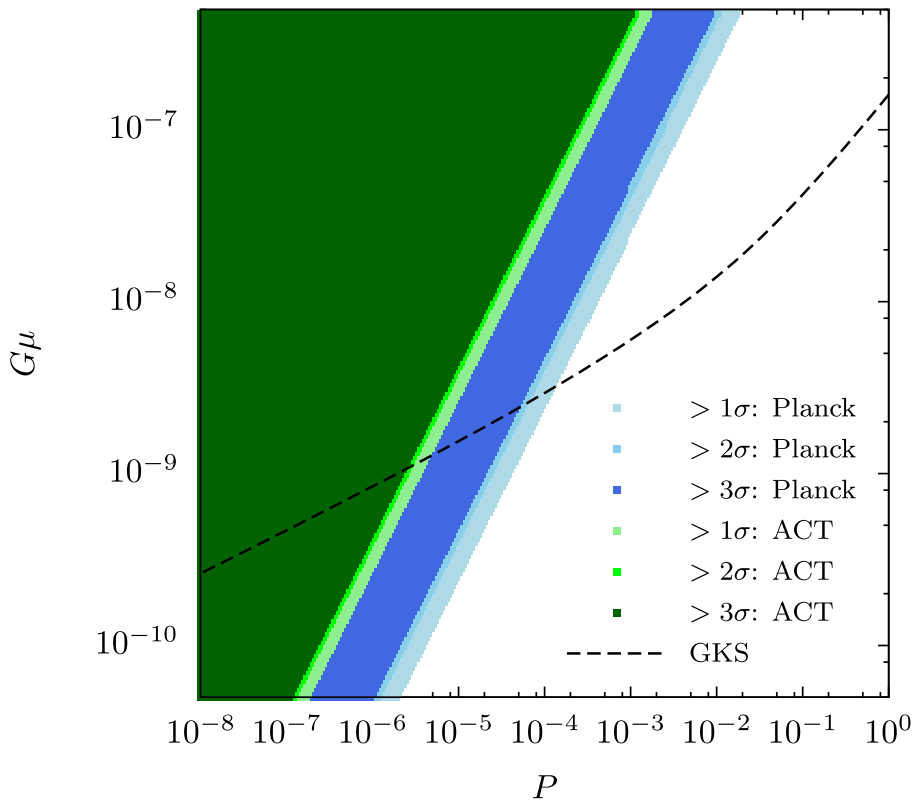}
\caption{
Constraints on string parameters, $P$ and $G\mu$, with the measured curl-mode power spectra. 
The colored regions are excluded with $>3\sigma$, $>2\sigma$ and $>1\sigma$ statistical significance 
with the curl mode from Planck (blue colored) and ACT (green colored). 
The black dashed line shows the lower bound of 
the string parameters disfavored by the temperature angular power spectrum~\cite{Yamauchi:2010ms}.
}
\label{P-Gmu}
\ec
\end{figure*}
%<><><><><><><><><><><><><><><><><><><><><><><><><><><><><><><><><><><><><><><><><><><><><><><><><><>%

Now let us show the constraints on cosmic string parameters, i.e., string tension $G\mu$ and 
reconnection probability $P$, based on the measured curl-mode power spectrum obtained with the BHE.
We compute the likelihood on the two-dimensional parameter space, $G\mu$ and $P$, given by 
%----------------------------------------------------------------------------------------------------%
\al{
	-2\ln \mC{L} (G\mu,P) = \sum_{b=1}^n 
		\frac{(\hC^{\mC{B}\mC{B}}_b - C^{\mC{B}\mC{B},{\rm theo}}_b(G\mu,P))^2}{(\sigma^{\mC{B}}_b)^2} 
	\,, 
}
%----------------------------------------------------------------------------------------------------%
where the quantity, $C^{\mC{B}\mC{B},{\rm theo}}_b(G\mu,P)$, is the theoretical binned power spectrum 
computed with specific values of $G\mu$ and $P$, with the same binning as shown in the measured power 
spectrum. 

First, we focus on the ordinary field-theoretic strings with $P=1$.
Assuming $P=1$\,, we find the upper bound of the string tension from the curl-mode power spectrum 
from ACT and Planck as
%----------------------------------------------------------------------------------------------------%
\al{ 
	&G\mu\leq
		8.9\times 10^{-4}\ \ (\,2\sigma\,, P=1\,,{\rm ACT}\,)
	\,, \\
	&G\mu\leq 
		6.6\times 10^{-5}\ \ (\,2\sigma\,, P=1\,,{\rm Planck}\,)
	\,, 
} 
%----------------------------------------------------------------------------------------------------%
respectively. These are rather weaker constraints than those obtained from temperature anisotropies 
through the GKS effect \cite{Dunkley:2010ge,Ade:2013xla}. 
On the other hand, for small values of $P$, the constraint on $G\mu$ from curl modes 
becomes tighter compared to that from the GKS effect. 
In Fig.~\ref{P-Gmu}, the resultant likelihood contours are shown for ACT (green colored) and Planck 
(blue colored), respectively. 
As we stated in Sec.~\ref{sec2}, the curl mode gives constraints on the combination of the string 
parameters as $G\mu P^{-1}$. 
Using ACT curl-mode data, we obtain constraints on the combination as 
%----------------------------------------------------------------------------------------------------%
\al{
	G\mu P^{-1} \leq 3.2\times 10^{-4}\ \ (\,2\sigma\,,{\rm ACT}\,)
	\,,
} 
%----------------------------------------------------------------------------------------------------%
with $P\lsim 10^{-2}$. 
For Planck data this can be improved to
%----------------------------------------------------------------------------------------------------%
\al{ 
	G\mu P^{-1} \leq 3.4\times 10^{-5}\ \ (\,2\sigma\,,{\rm Planck}\,)
	\,.
} 
%----------------------------------------------------------------------------------------------------%
For comparison, the black solid line in Fig.~\ref{P-Gmu} represents the lower bound of the string 
parameters disfavored by the temperature power spectrum \cite{Yamauchi:2010ms}. 
For $P\lsim 10^{-4}$ (\,$P\lsim 5\times 10^{-6}$\,), the constraint using the curl-mode power spectrum 
from Planck (ACT) is tighter than that from the GKS-induced temperature power spectrum.
The curl mode is also useful in terms of robust constraints on cosmic strings, because the constraints 
from the GKS effect rely entirely on the measurement of the temperature power spectrum at small scales 
where the contributions from other secondary effects such as point sources and the SZ effect are 
usually significant. 
Note that the curl-mode power spectrum has been also measured from the South Pole Telescope 
temperature map \cite{vanEngelen:2012va}, and compared to the ACT data, it gives a higher statistical 
significance on small scales. 
However, we expect that the constraints on $G\mu$ and $P$ from SPT are not dramatically improved so 
much, since these constraints mainly come from the lower-multipole power, for which we do not find any 
big difference between the ACT and SPT data.

Note that the error bars for ACT are estimated in an idealistic case, and the parameter constraints  
here may be underestimated. 
Even in this case, if the constraints were inflated by a factor of 2, the uncertainties in the model 
of cosmic strings would still be large. 
Our primary purpose is to show an example of cosmological application of curl mode. 
In this sense, although the constraints obtained here may be degraded, the curl mode is still useful 
to constrain cosmic strings.

%% file: ms-sec6.tex
%****************************************************************************************************%
%////////////////////////////////////////////////////////////////////////////////////////////////////%
\section{Summary and future prospects} 
\label{sec6}
%////////////////////////////////////////////////////////////////////////////////////////////////////%
%****************************************************************************************************%

%<><><><><><><><><><><><><><><><><><><><><><><><><><><><><><><><><><><><><><><><><><><><><><><><><><>%
\begin{figure*}[t]
\bc
\includegraphics[width=100mm,clip]{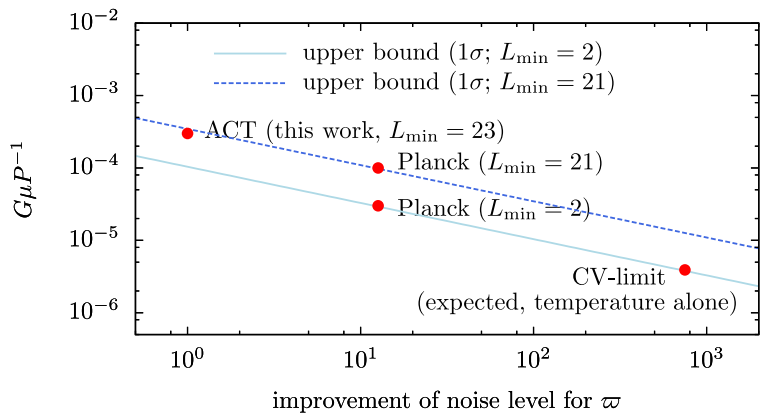} 
\caption{
The $1\sigma$ constraints on the combination of cosmic string parameters, $G\mu P^{-1}$, as a function 
of improvement factor for the signal-to-noise ratio of curl mode normalized by the ACT case. 
For Planck, since the parameter constraint highly depends on the minimum value of curl mode multipoles 
for parameter constraints, $L\rom{min}$, we also show the case with $L\rom{min}=21$. 
The two lines show the expected upper bound on $G\mu P^{-1}$, with $L\rom{min}=2$ (solid) or 
$L\rom{min}=21$ (dashed), respectively. 
For comparison, we also show the case with cosmic-variance limit (CV-limit) with $L\rom{min}=2$, where 
we assume that the temperature multipoles up to $\ell=7000$ are used for lensing reconstruction. 
}
\label{Fig:Planck}
\ec
\end{figure*}
%<><><><><><><><><><><><><><><><><><><><><><><><><><><><><><><><><><><><><><><><><><><><><><><><><><>%

% Summary of our results ----------------------------------------------------------------------------%
In this paper, we have presented constraints on cosmic strings with the curl-mode power spectrum from 
ACT and Planck. 
We first show the gradient and curl mode measured from ACT temperature map, and show that both the 
parameters, $\Ag$ and $\Ac$, are consistent between different methods and with other data within 
$1\sigma$ statistical significance, implying that residual systematics should be negligible in the 
resultant constraints on cosmic string parameters. 
Based on the measured curl-mode from ACT and Planck, we then obtain constraints on $P$ and $G\mu$. 
Although, for $P=1$, the constraint on $G\mu$ is weaker compared to the current constraint from the 
temperature power spectrum, we found that the constraints on the string parameters with the 
reconnection probability $P\lsim 10^{-4}$ become tighter than those from the temperature power spectrum 
via the GKS effect. 
With Planck data, we obtained an upper bound on the combination of the string parameters as
$G\mu P^{-1} \lsim 3.4\times 10^{-5}$ with $P\lsim 10^{-2}$. 

Note here that measurement of the curl-mode power spectrum can provide stringent constraints on the 
properties of the cosmic superstring network with an extremely small value of $P$.
For instance, it is interesting to investigate a hybrid network model that contains bound states, in 
which strings with various values of the reconnection probability can be formed. 
The resultant curl-mode power spectrum generated from such network would be not so different from that 
used in this paper, and the qualitative constraints on a hybrid network could be also similar to those 
obtained in our analysis. 
The complete derivation of the curl-mode power spectrum from such network model is interesting, 
and we hope to come back to this issue in a future publication.

% future prospects ----------------------------------------------------------------------------------%
We now turn to discuss how the constraint on cosmic string parameters will be improved by the quality 
of curl-mode measurements. 
In Fig.~\ref{Fig:Planck}, we show the constraint on $G\mu P^{-1}$ as a function of the improvement 
factor of the noise level of curl mode. 
Here we assume that only temperature maps are used for lensing reconstruction. For the Planck case, we 
also show results with the minimum value of curl-mode multipoles as $L\rom{min}=21$. 
The result implies that, a measurement of lower multipoles of the curl-mode power spectrum is 
essential to constrain the string parameters. 
We also show the expected upper bound on $G\mu P^{-1}$ as a function of the improvement factor, 
extrapolated from the Planck results, in the case with $L\rom{min}=2$ (solid line) and $L\rom{min}=21$ 
(dashed line). 
Fig.~\ref{Fig:Planck}, apparently indicates that the case of cosmic variance limit is not so improved 
compared to that of Planck. 
This implication, however, does not limit the potential of curl mode for constraining a model of 
cosmic string and/or other vector and tensor sources. 
In near future, several ground-based CMB experiments will focus on a precise measurement of B-mode 
polarization. 
In the case of the cosmic-variance limit, the inclusion of polarization improves the 
signal-to-noise ratio of curl mode by $2$ orders of magnitude compared to the case without 
polarization \cite{Namikawa:2011cs}. 
This leads to an expected constraint on $G\mu P^{-1}$ of $\sim 5\times 10^{-7}$. 
Moreover, as shown in Ref.~\cite{Hirata:2003ka}, measurements of B-mode polarization on small scales 
directly probe the lensing potentials, and thus the Gaussian bias, in principle, vanishes in the 
absence of primordial B-mode polarizations and other non-lensing B-mode sources. 
Therefore, measurements of polarization anisotropies should eventually lead to a high-precision 
lensing reconstruction, and this can give a further improvement of constraints on cosmic string 
parameters.

%% file: ms-appA.tex
%****************************************************************************************************%
%////////////////////////////////////////////////////////////////////////////////////////////////////%
\section{Error estimate of angular power spectrum} 
\label{appA} 
%////////////////////////////////////////////////////////////////////////////////////////////////////%
%****************************************************************************************************%

%<><><><><><><><><><><><><><><><><><><><><><><><><><><><><><><><><><><><><><><><><><><><><><><><><><>%
\begin{table}[t]
\caption{
\label{lmin-lmax}
Constraints on $\Ag$ while varying the maximum value of temperature multipoles used for lensing 
reconstruction, $\ell\rom{max}$. 
}
\vs{0.5}
\begin{tabular}{c|ccc} \hline 
$\ell\rom{max}$ & $1900$ & $2000$ & $2100$ \\ \hline 
$\Ag$ & $1.00 \pm 0.51$ & $0.91\pm 0.46$ & $0.88\pm 0.44$ \\ \hline 
$\Ac$ & $<0.91$ & $<0.80$ & $<0.79$ \\ \hline 
\end{tabular}
\end{table}
%<><><><><><><><><><><><><><><><><><><><><><><><><><><><><><><><><><><><><><><><><><><><><><><><><><>%

%<><><><><><><><><><><><><><><><><><><><><><><><><><><><><><><><><><><><><><><><><><><><><><><><><><>%
\begin{table}[t]
\caption{
\label{binning}
Same as Table \ref{lmin-lmax}, but for a varying number of lensing multipoles bins, $n$. 
}
\begin{tabular}{c|ccc} \hline 
Number of bins & $7$ & $8$ & $9$ \\ \hline 
$\Ag$ & $0.91\pm 0.46$ & $0.91\pm 0.46$ & $0.91\pm 0.46$ \\ \hline 
$\Ac$ & $<0.81$ & $<0.80$ & $<0.86$ \\ \hline 
\end{tabular}
\vs{0.5}
\end{table}
%<><><><><><><><><><><><><><><><><><><><><><><><><><><><><><><><><><><><><><><><><><><><><><><><><><>%

%<><><><><><><><><><><><><><><><><><><><><><><><><><><><><><><><><><><><><><><><><><><><><><><><><><>%
\begin{table}[t]
\caption{
\label{ell-splitting}
Same as Table \ref{lmin-lmax}, but for comparing with the results obtained by $\ell$-splitting. 
The multipole range of $\ell$-splitting is varied as 
$([500,1500],[1600,2300])$ (LSP), 
$([500,1550],[1600,2300])$ (LSP$^{\prime}$), 
$([500,1500],[1550,2300])$ (LSP$^{\prime\prime}$)
. 
}
\vs{0.5}
\begin{tabular}{c|cccc} \hline 
Method & BHE & LSP & LSP$^{\prime}$ & LSP$^{\prime\prime}$ \\ \hline 
$\Ag$ & $0.91\pm 0.46$  & $1.06 \pm 0.66$ & $1.01\pm 0.63$ & $1.26\pm 0.66$ \\ \hline 
$\Ac$ & $<0.80$ & $< 0.97$ & $< 0.98$ & $< 0.93$ \\ \hline 
\end{tabular}
\end{table}
%<><><><><><><><><><><><><><><><><><><><><><><><><><><><><><><><><><><><><><><><><><><><><><><><><><>%

%++++++++++++++++++++++++++++++++++++++++++++++++++++++++++++++++++++++++++++++++++++++++++++++++++++%
%----------------------------------------------------------------------------------------------------%
\subsection{Expected error estimate} 
%----------------------------------------------------------------------------------------------------%
%++++++++++++++++++++++++++++++++++++++++++++++++++++++++++++++++++++++++++++++++++++++++++++++++++++%

For the $i$th map, given Fourier multipoles, $X_{\bL}$, the optimal unbiased estimator for the angular 
power spectrum is given by 
%----------------------------------------------------------------------------------------------------%
\al{
	\hC^{XX}_{L} = \frac{1}{W N_{L}}\sum_{|\bL|=L} |X_{\bL}|^2 
	\,, 
}
%----------------------------------------------------------------------------------------------------%
where $N_{L}$ is the number of fluctuations whose multipole coefficient satisfies $|\bL|=L$, and we 
assume 
%----------------------------------------------------------------------------------------------------%
\al{
	\ave{X_{\bL}X_{\bL'}} = W C^{XX}_{L}\delta_{\bL-\bL'}
	\,. 
}
%----------------------------------------------------------------------------------------------------%
The normalization, $W$, is usually arising from, e.g., the effect of a window function. 
The error is then estimated as 
%----------------------------------------------------------------------------------------------------%
\al{
	[\Delta C_{L}^{XX}]^2 
		&= \frac{1}{N_L^2}\sum_{|\bL|=L}\sum_{|\bL'|=L} \frac{\ave{|X_{\bL}|^2|X_{\bL'}|^2}}{W^2} 
		- (C^{XX}_{L})^2
	\notag \\ 
		&= \frac{1}{N_L^2}\sum_{|\bL|=L}\frac{2(W C_L^{XX})^2}{W^2} 
	\notag \\ 
		&= \frac{2}{N_{L}}(C_{L}^{XX})^2 
	\,, \label{Eq:error}
}
%----------------------------------------------------------------------------------------------------%
where we assume $X_{\bL}$ is a random Gaussian field. 
We will use the above expression for the error estimate. 

%++++++++++++++++++++++++++++++++++++++++++++++++++++++++++++++++++++++++++++++++++++++++++++++++++++%
%----------------------------------------------------------------------------------------------------%
\subsection{Map-combined power spectrum} 
%----------------------------------------------------------------------------------------------------%
%++++++++++++++++++++++++++++++++++++++++++++++++++++++++++++++++++++++++++++++++++++++++++++++++++++%

In our analysis, the angular power spectra, $C_L^{\mC{E}\mC{E}}$ and $C_L^{\mC{B}\mC{B}}$, are first 
estimated in each map based on Eq.~(\ref{Eq:est-Clxx}). 
The map-combined power spectrum is then obtained by weighting the inverse of its variance 
(e.g., \cite{vanEngelen:2012va}): 
%----------------------------------------------------------------------------------------------------%
\al{
	\widehat{C}^{XX}_{L} 
		= \left\{\sum_{i} \frac{1}{[\Delta C^{XX}_{L,i}]^2}\right\}^{-1} 
		\sum_{i} \frac{\widehat{C}^{XX}_{L,i}}{[\Delta C^{XX}_{L,i}]^2}
	\,, 
}
%----------------------------------------------------------------------------------------------------%
where $X=\mC{E}$ or $\mC{B}$, and the variance of the angular power spectrum in each map is estimated 
with the ideal case based on Eq.~(\ref{Eq:error}): 
%----------------------------------------------------------------------------------------------------%
\al{
	[\Delta C_{L,i}^{XX}]^2 
		&= \frac{2}{N_{L}}(\mC{A}_{L,i}^{XX}+C_{L}^{XX})^2 
	\,, \label{Eq:est-error}
}
%----------------------------------------------------------------------------------------------------%
with $\mC{A}_{L,i}^{XX}$ denoting the normalization (\ref{Eq:mcA}) for the $i$th map, and $N_L$ being 
the number of fluctuations whose multipole coefficient satisfies $|\bL|=L$. 
For $\ell$-splitting, instead of Eq.~(\ref{Eq:est-error}), we estimate the error of the measured power 
spectrum as 
%----------------------------------------------------------------------------------------------------%
\al{
	[\Delta C_{L,i}^{XX}]^2 = \frac{1}{N_{L}} 
		(\mC{A}_{L,i}^{XX,(1)}+C_{L}^{XX,{\rm fid}})(\mC{A}_{L,i}^{XX,(2)}+C_{L}^{XX,{\rm fid}}) 
	\,, \label{Eq:est-error-lsp}
}
%----------------------------------------------------------------------------------------------------%
where the quantities, $\mC{A}_{L,i}^{XX,(1)}$ and $\mC{A}_{L,i}^{XX,(2)}$, are the normalization 
computed on each disjoint annulus.

%% file: ms-appB.tex
%****************************************************************************************************%
%::::::::::::::::::::::::::::::::::::::::::::::::::::::::::::::::::::::::::::::::::::::::::::::::::::%
\section{Test for systematic uncertainties} 
\label{appB}
%::::::::::::::::::::::::::::::::::::::::::::::::::::::::::::::::::::::::::::::::::::::::::::::::::::%
%****************************************************************************************************%

In this section, we show several tests for systematic uncertainties in estimating lensing power 
spectra.

%++++++++++++++++++++++++++++++++++++++++++++++++++++++++++++++++++++++++++++++++++++++++++++++++++++%
%----------------------------------------------------------------------------------------------------%
\subsubsection{Temperature multipoles, $\ell\rom{min}$ and $\ell\rom{max}$} 
%----------------------------------------------------------------------------------------------------%
%++++++++++++++++++++++++++++++++++++++++++++++++++++++++++++++++++++++++++++++++++++++++++++++++++++%

Here we show the dependence of our results on the range of temperature multipoles. 
The results of parameter constraints on $\Ag$ and $\Ac$ are given in Table \ref{lmin-lmax}.

%++++++++++++++++++++++++++++++++++++++++++++++++++++++++++++++++++++++++++++++++++++++++++++++++++++%
%----------------------------------------------------------------------------------------------------%
\subsubsection{Binning of measured power spectrum} 
%----------------------------------------------------------------------------------------------------%
%++++++++++++++++++++++++++++++++++++++++++++++++++++++++++++++++++++++++++++++++++++++++++++++++++++%

To test whether or not our results depend on the binning of measured lensing power spectra, 
we compute the constraints on $\Ag$ and $\Ac$, while varying the number of multipole bins, $n$. 
In Table \ref{binning}, to compare with our fiducial number of bins, $n=8$, we show $n=8\pm 1$ cases.

%++++++++++++++++++++++++++++++++++++++++++++++++++++++++++++++++++++++++++++++++++++++++++++++++++++%
%----------------------------------------------------------------------------------------------------%
\subsubsection{Method of lensing reconstruction} 
%----------------------------------------------------------------------------------------------------%
%++++++++++++++++++++++++++++++++++++++++++++++++++++++++++++++++++++++++++++++++++++++++++++++++++++%

To test the effect of these biases, we compare with the results of lensing reconstruction with the 
$\ell$-splitting method. 
In Table \ref{ell-splitting}, we show the results of constraints on $\Ag$ and $\Ac$, while varying 
several cases of two disjoint annuli.